\documentclass[a4paper,UKenglish,cleveref, autoref, thm-restate, nolinenumbers]{lipics-v2021}

\usepackage{mathrsfs}
\usepackage[T1]{fontenc}
\usepackage{graphicx}
\usepackage[utf8]{inputenc}
\usepackage{newunicodechar}
\newunicodechar{ℛ}{\mathcal{R}}
\usepackage{multirow}
\usepackage{amsmath}
\usepackage{listings}
\usepackage{longtable}
\usepackage{tabularx}
\usepackage{array}
\usepackage{tabularray}
\usepackage{stmaryrd}
\usepackage{subcaption}

\hideLIPIcs 

\title{The \texorpdfstring{$\mathbf{P}$}{P}-Completeness of Inverted Index Traversal: On the Complexity of Evaluating Boolean Query DAGs}

\author{Amir Aavani}{Apple Inc., USA}{aaavani@apple.com}{}{}
\authorrunning{A. Aavani}
\Copyright{Amir Aavani}


\category{} 
\relatedversion{} 
\supplement{}
\funding{}

\begin{document}
\nolinenumbers  

\maketitle

\begin{abstract}
Modern AI agents increasingly rely on search infrastructure to execute complex, neuro-symbolic reasoning workflows. These workflows often compile into deeply nested, non-monotonic Boolean queries over text fields. However, standard query evaluation strategies over inverted indices face structural limitations when handling these topologies. We establish that stateful iterator models (Document-at-a-Time) are bounded by $\text{NC}^1$ formula evaluation. Consequently, any polynomial-time query rewrite mechanism attempting to evaluate re-convergent logic on these engines is mathematically forced to produce an execution plan of super-polynomial size, resulting in an $O(2^{|Q|})$ exponential increase in time complexity. Conversely, recursive materialization models (Term-at-a-Time) incur an $\Omega(|U|)$ space complexity requirement when evaluating logical negation over the document universe.

In this paper, we analyze the theoretical boundaries of executing complex logic natively over an inverted index. We formalize a retrieval language ($\mathcal{L}_R$) based on Directed Acyclic Graphs (DAGs) and prove that its evaluation problem is  \textbf{$\mathbf{P}$-Complete}. To evaluate this language efficiently, we introduce \texttt{ComputePN}, a deterministic, sparsity-aware evaluation algorithm. By decoupling logical negation from universe-proportional materialization via a Positive-Negative dual representation, and utilizing native DAG memoization over partitioned data blocks, \texttt{ComputePN} evaluates $\mathbf{P}$-Complete queries with a time complexity of $O(|Q| \cdot |U_{\mathit{active}}|)$ and an auxiliary space complexity of $\mathcal{O}(M)$ for any memory limit $M \ge |Q|$. This approach avoids both the exponential tree-expansion overhead and the universal materialization requirement, providing a formal foundation for computational retrieval.
\end{abstract}

\maketitle

\section{Introduction}
\label{secIntroduction}

The inverted index is the foundational data structure for large-scale information retrieval (IR) systems~\cite{witten1999managing,croft2010search}. Beyond dedicated search engines, it serves as a critical execution component in modern relational databases---such as PostgreSQL's Generalized Inverted Index (GIN)~\cite{obart2014postgresql}---to accelerate queries over text, arrays, and semi-structured data. 

Traditional query evaluation over documents has evolved into hybrid retrieval architectures. Modern search engines typically execute a dual-path first stage: employing Approximate Nearest Neighbor (ANN) indices for dense semantic candidate generation~\cite{karpukhin2020dense}, operating in parallel with an inverted index for sparse lexical matching. The retrieved candidates are subsequently processed by a computationally intensive downstream scoring or re-ranking phase (e.g., Learning-to-Rank models or Cross-Encoders) to finalize the results~\cite{wang2011cascade, gao2022neural}.

However, while dense retrieval excels at semantic approximation, fixed-dimension embedding spaces are mathematically incapable of reliably enforcing strict logical exclusions or non-monotonic boundaries~\cite{weller2025theoretical}. Consequently, across both modern hybrid search platforms and relational databases, the inverted index remains the mandatory execution layer for guaranteeing the deterministic evaluation of Boolean constraints.

To optimize this strict Boolean evaluation phase, query execution engines typically employ one of two primary strategies: 
\begin{itemize}
    \item \textbf{Materialized Evaluation (Term-at-a-Time / TAAT):} Operating operator-by-operator, this strategy fully evaluates one query node at a time, materializing intermediate results as sets or compressed bitmaps before passing them to the next operation~\cite{moffat1996self}.
    \item \textbf{Pipelined Evaluation (Document-at-a-Time / DAAT):} Formalized in early retrieval literature~\cite{turtle1995query}, this strategy evaluates queries by concurrently advancing stateful iterators across the posting lists. Conceptually isomorphic to the Volcano iterator model in relational databases~\cite{graefe1993encapsulation}, DAAT avoids intermediate materialization by resolving the Boolean logic dynamically as the iterators align on matching document identifiers.
\end{itemize}

Both strategies are highly optimized for their historical workload: monotonic top-$k$ evaluation over sparse text fields.

\subsection{The Complexity Bottleneck of Modern Workloads}
As we detail in Section~\ref{sec:resurgence}, the proliferation of Large Language Models (LLMs) and autonomous agents has introduced a new paradigm of information access. These neuro-symbolic workflows require the inverted index to evaluate massive, non-monotonic logical constraints. Mathematically, these constraints can be expressed in the form of \textbf{Directed Acyclic Graphs (DAGs)} of Boolean operators, where shared sub-computations are referenced by multiple parent nodes in the execution plan.

Standard query evaluation strategies face fundamental theoretical limitations when forced to execute these DAG-structured, non-monotonic constraints:

\begin{enumerate}
    \item \textbf{Query Complexity (The Factorization Failure):} Pipelined iterator models (DAAT) rely on stateful pointers that cannot branch or share state across an execution graph. Consequently, to evaluate a DAG purely via pipelining, the execution plan must be ``unrolled'' into a strict syntactic tree. In computational complexity, evaluating a strict tree of Boolean operators is structurally equivalent to the \textbf{Boolean Formula Value Problem}, bounding the execution model to the complexity class $\text{NC}^1$. As we prove in Theorem~\ref{thm:daat_rewrite_lower}, there exists a strict family of queries for which any polynomial-time query rewrite mechanism attempting to map them onto DAAT iterators is mathematically forced to produce an execution plan of super-polynomial size, resulting in a worst-case $O(2^{|Q|})$ exponential expansion. If the system attempts to avoid this expansion by caching the shared nodes, it abandons the pipelined DAAT paradigm entirely, reverting to the materialization overhead of TAAT.

    \item \textbf{Data Complexity (The Universal Scan):} Materialized evaluation models (TAAT) naturally support DAGs, as intermediate variables can be cached and reused. However, they suffer a critical failure in \textit{data complexity} when evaluating logical negation ($\neg$). Evaluating an unrestricted negation over an inverted index requires materializing the complement of a posting list relative to the universe of documents $U$. This forces the engine to allocate $\Omega(|U|)$ memory and processing time. In web-scale databases where $|U|$ is large and active posting lists are sparse, this ``Universal Scan'' requirement renders evaluation intractable.
\end{enumerate}

Evaluating strict, non-monotonic logical DAGs over inverted indices without triggering universe-proportional materialization or exponential tree unrolling remains a fundamental theoretical challenge. To scale agentic search, we must bridge this gap with a \textbf{sparsity-aware} execution engine capable of natively evaluating arbitrary tractable constraints.

\subsection{Contributions}
In this paper, we analyze the theoretical boundaries of executing complex logic natively over an inverted index. We frame the index not merely as a sparse filter, but as a generalized computational processor. Our contributions are:

\begin{itemize}
    \item \textbf{The $\mathbf{P}$-Completeness of DAG Retrieval:} We define a formal query language ($\mathcal{L}_R$) based on Boolean DAGs, which is descriptively equivalent to the class of polynomial-size propositional circuits. Connecting to the foundations of computational complexity, we provide a constructive LogSpace reduction from the Circuit Value Problem to prove that the decision problem for evaluating $\mathcal{L}_R$ over an inverted index is \textbf{$\mathbf{P}$-Complete}.

    \item \textbf{The Query Rewrite Lower Bound:} We establish a formal gap in evaluation complexity between traditional DAAT iterators ($\text{NC}^1$) and neuro-symbolic workloads ($\mathbf{P}$). We prove that there exists a family of $\mathbf{P}$-Complete DAGs for which any polynomial-time query rewrite mechanism attempting to map them onto stateful iterators is mathematically forced to produce an execution plan of super-polynomial size, resulting in a worst-case $O(2^{|Q|})$ exponential expansion.
    
    \item \textbf{Sparsity-Aware Evaluation (\texttt{ComputePN}):} While the $\mathbf{P}$-Completeness of evaluating Boolean DAGs is a straightforward consequence of circuit complexity, executing these queries over massive indices without triggering universe-proportional materialization on negations remains a fundamental algorithmic challenge. We introduce \texttt{ComputePN}, a deterministic evaluation algorithm that addresses this. By adapting symbolic complement techniques into a ``Positive-Negative'' set-algebra, \texttt{ComputePN} achieves an execution time of $O(|Q| \cdot |U_{\text{active}}|)$. This preserves sparsity, avoiding both the exponential tree-expansion overhead and the universal scan requirement.
    
    \item \textbf{Validating the Neuro-Symbolic QEP:} We demonstrate the feasibility of our theoretical bounds by treating an LLM as a heuristic query compiler. We show that complex arithmetic constraints can be compiled into Query Execution Plans (QEPs) of hundreds of nodes. While legacy iterators exceed structural limits ($>10,000$ clauses), \texttt{ComputePN} evaluates the equivalent topological circuits natively over an 8.8M document index, proving the practicality of polynomial-time boolean execution.
\end{itemize}
\section{The Resurgence of Boolean Retrieval}
\label{sec:resurgence}

For the past two decades, the Information Retrieval (IR) community has largely shifted focus away from strict Boolean execution, prioritizing soft-matching and ranking (e.g., BM25, Learning-to-Rank, and Dense Vectors). This shift was driven by human behavior: users typically issue short, ambiguous keyword queries, making strict logical filtering brittle and prone to zero-recall failures. Consequently, modern query execution engines are highly optimized for Top-$k$ ranking heuristics rather than deep logical evaluation.

However, the proliferation of Large Language Models (LLMs) and autonomous agents (e.g., RAG~\cite{lewis2020retrieval}) has introduced a new paradigm of information access. In modern neuro-symbolic workflows, the entity querying the index is no longer a human typing keywords, but an LLM acting as a query compiler. When tasked with complex reasoning, the LLM may need to translate the user's intent into massive, precise, non-monotonic constraints. 

\subsection{The Failure of Fetch-and-Filter}
To evaluate these complex constraints, modern agentic frameworks typically rely on a ``Fetch-and-Filter'' pattern. The agent issues a broad, semantic query to the retrieval engine, fetches the Top-$k$ results, and then manually evaluates the strict logical constraints (e.g., date boundaries, exclusivity, or arithmetic thresholds) within the application layer or the LLM's context window.

This approach suffers from the \textbf{Recall Failure}. If the logical constraints are highly restrictive, the few valid documents may be buried deep within the semantic long-tail (e.g., ranked at position 50,000). Because the agent can only fetch and process a limited number of documents due to network latency and context-window limits, there is always the risk of missing some valid documents. 

To guarantee correct execution, the strict logical computation must be pushed down to the \textbf{inverted index}. The retriever must act as a deterministic logic processor, evaluating the Boolean constraints natively across the entire universe of documents before any ranking or truncation occurs.

\subsection{Motivational Examples}
To illustrate the necessity of pushing complex Boolean logic down to the inverted index, consider the following neuro-symbolic agentic workflows. In each case, the LLM must enforce a strict logical constraint to guarantee the correctness of the retrieved context.

\noindent \textbf{1. Concept Disentanglement (Deep Negation):} 
Consider a research agent tasked with finding foundational literature on a specific algorithm, but strictly excluding modern neural implementations or non-peer-reviewed preprints. 

\textit{User Intent:} ``Find papers discussing `Gradient Descent', but exclude any papers tagged with \texttt{Venue:ArXiv} or containing the keyword `Deep Learning'.''

To enforce this, the LLM must generate a query with deep, disjunctive negation: $\texttt{Gradient Descent} \land \neg (\texttt{Venue:ArXiv} \lor \texttt{Deep Learning})$. If the engine relies on semantic soft-matching, the high latent proximity between these concepts often causes ``Semantic Bleed,'' where negated concepts are accidentally retrieved due to contextual similarity. Strict Boolean negation over deterministic fields is required to guarantee the integrity of the agent's context window.

\noindent \textbf{2. Dynamic Aggregation (Re-convergent Logic):} 
Consider a bibliometric research agent evaluating the aggregate impact of a paper's authors.

\textit{User Intent:} ``Find papers where the combined citation count of the paper's authors exceeds 100,000.''

Because the set of authors changes per document, this mathematical inequality must be evaluated dynamically over the document's attributes. If the agent relies on a ``Fetch-and-Filter'' pattern, it must issue a broad query (e.g., fetching all papers by highly cited authors), load thousands of documents over the network into the application layer, and manually compute the sums. If the threshold is highly restrictive, the few valid papers may be buried deep in the retrieval tail, causing a catastrophic recall failure when the agent hits its context or pagination limits. 

To guarantee correct execution without network exhaustion, this arithmetic constraint must be pushed down into the inverted index as a Boolean circuit over bit-sliced posting lists. This arithmetic logic is inherently re-convergent, meaning intermediate sub-computations (such as the presence of a specific author) are shared across multiple branches of the logic gates to compute the final sum.

\vspace{1em}
While BM25 and other token-based scoring mechanisms are still required for the final ranking phase, the retriever---specifically the inverted index---must first act as a deterministic logic processor to enforce the hard boundaries of these constraints. 

As we will formally demonstrate in Section~\ref{sec:limits}, when LLMs compile high-level reasoning into strict Boolean constraints, they naturally generate highly re-convergent, non-monotonic logic structures. Existing retrieval architectures and relational databases are mathematically unequipped to execute these structures efficiently, forcing a strict dichotomy between exponential execution time and universe-scale memory exhaustion.

\section{Related Work}
\label{sec:related}

Modern information retrieval architectures can be broadly categorized into two primary paradigms: \textbf{Token-based} (Sparse) retrieval and \textbf{Embedding-based} (Dense) retrieval. 

To understand the structural limits analyzed in Section~\ref{sec:limits}, we must first examine the formal execution mechanics of these standard architectures. We evaluate them specifically on their ability to execute the \textbf{full spectrum of Boolean logic} ($\land, \lor, \neg$) structured as \textbf{Directed Acyclic Graphs (DAGs)}. As established in Section~\ref{sec:language}, supporting this topological structure is a fundamental requirement for achieving $\mathbf{P}$-Complete query evaluation.

\subsection{Term-at-a-Time (TAAT) and Space Complexity}
In the TAAT strategy (Token-based), the engine processes posting lists one by one, fully evaluating one query operator before passing the materialized result to the next~\cite{moffat1996self,witten1999managing}.

\begin{itemize}
    \item \textbf{Mechanism:} Intermediate results are materialized as sets or compressed bitmaps (e.g., Roaring Bitmaps~\cite{chambi2016better}) and logically combined. This structure naturally supports \textbf{DAGs}, as the result of any shared sub-expression can be cached in memory and reused by multiple parent nodes.
    
    \item \textbf{The Sparsity Barrier:} While TAAT bounds query complexity, it suffers a critical failure in \textit{Space Complexity} when evaluating non-monotonic queries. Specifically, evaluating an unrestricted logical negation ($\neg A$) requires materializing the complement of $A$ relative to the universe of documents $U$. Because TAAT evaluates queries strictly bottom-up, it explicitly materialize this dense complement before passing it to subsequent operators. Even if the engine utilizes compressed bitmaps, the complement of a highly sparse posting list yields a massively dense, fragmented bitset that resists efficient compression. Consequently, evaluating these non-monotonic nodes forces the engine to allocate auxiliary space proportional to the universe size ($\Omega(|U|)$), violating the principle of \textbf{Sparsity-Aware Evaluation} and rendering TAAT practically unscalable for web-scale indices.
\end{itemize}

\subsection{Document-at-a-Time (DAAT) and \texorpdfstring{$\text{NC}^1$}{NC1} Limits}
The DAAT strategy, employed by modern engines like Lucene~\cite{bialecki2012apache}, processes posting lists concurrently within a single pass using a hierarchical tree of iterators~\cite{moffat1996self}.

\begin{itemize}
    \item \textbf{Mechanism:} The engine constructs an \textbf{Iterator Tree} mirroring the query's syntactic structure. Rather than materializing intermediate sets, internal nodes advance their child iterators in a coordinated, interleaved fashion. They utilize \textbf{ZigZag joins} (to intersect streams via skip pointers) or \textbf{Min-Heaps} (to merge streams). This allows the engine to evaluate the logic document-by-document, identifying candidates in $O(1)$ auxiliary space.
    
    \item \textbf{The WAND Limitation:} Dynamic pruning algorithms like WAND~\cite{broder2003efficient} extend DAAT to accelerate Top-$k$ ranking. However, this optimization is \textbf{ineffective} for strict Boolean retrieval. WAND relies on monotonic score variance to prune documents that cannot exceed a threshold. In a strict Boolean context, sub-expressions produce uniform binary outcomes ($0$ or $1$) and the retrieval goal is to identify \textit{all} valid assignments (Recall $= 1.0$) rather than a ranked Top-$k$. This renders the WAND pruning mechanism degenerate. Furthermore, WAND provides no mechanism to avoid the linear scan required for \textbf{Disjunctive Negation} ($A \lor \neg B$), as it fundamentally assumes terms contribute positively to a cumulative score.
    
    \item \textbf{The $\text{NC}^1$ Ceiling:} Because DAAT iterators are stateful (encapsulating a specific document pointer), a single iterator instance cannot be shared across multiple branches of the execution graph. Consequently, the execution plan is isomorphic to a Boolean Formula. From a descriptive complexity standpoint, this bounds the query evaluation complexity of DAAT to the \textbf{Boolean Formula Value Problem}, which is complete for $\text{NC}^1$. To evaluate a DAG (which corresponds to the \textbf{Circuit Value Problem}, a language whose evaluation is $\mathbf{P}$-Complete), DAAT must unroll the DAG into an $\text{NC}^1$ formula. Simulating a Boolean circuit with a Boolean formula requires a worst-case $O(2^{|Q|})$ exponential blowup in query size.
    
    \item \textbf{Graph-Structured IR Languages (e.g., Galago):} Some academic retrieval frameworks, such as Galago~\cite{croft2010search}, introduce graph-based query languages to express complex dependencies. However, these frameworks primarily use the graph topology for \textit{query parsing and feature extraction} (e.g., computing term proximity or BM25 scores) rather than for strict, polynomial-time Boolean evaluation. When forced to execute strict logical intersections, these engines still rely on underlying DAAT iterators, meaning they either fall back to tree-expansion or resort to heavy intermediate materialization (TAAT) to resolve shared nodes, failing to achieve $O(|Q|)$ evaluation without memory overhead.
\end{itemize}

\subsection{Alternative Paradigms: Neural and Relational Engines}
While \textbf{Dense Retrieval} (Bi-Encoders) excels at semantic approximation~\cite{karpukhin2020dense}, it suffers from a fundamental Expressiveness Gap: fixed-dimension vectors cannot linearly separate complex Boolean concepts like parity or strict exclusion~\cite{weller2025theoretical}. Conversely, \textbf{Neural Re-rankers} (Cross-Encoders) process the query and document jointly through deep self-attention layers~\cite{nogueira2019passage}. While Transformers possess the theoretical expressive power to evaluate complex Boolean logic, they require $O(|U|)$ heavy inference steps, violating sparsity-aware evaluation bounds and rendering them intractable for corpus-wide logical filtering.

Furthermore, \textbf{Relational Databases} (e.g., PostgreSQL) support text retrieval via specialized inverted index structures like the Generalized Inverted Index (GIN)~\cite{obart2014postgresql}. While relational engines support Turing-complete logic via Recursive Common Table Expressions (CTEs), applying these formalisms to evaluate Boolean DAGs introduces a severe Impedance Mismatch. When standard SQL is used to query a GIN index, the query planner parses the \texttt{WHERE} clause into a syntactic execution tree. Like DAAT, this tree-based planner is structurally bounded by $\text{NC}^1$. If a complex, re-convergent DAG is expressed as a flat SQL query, the planner cannot natively factorize the shared sub-computations, leading to redundant bitset evaluations and exponential unrolling. While advanced relational techniques such as Worst-Case Optimal Joins (WCOJ)~\cite{ngo2018worst} achieve optimal bounds for multi-way intersections, they are structurally restricted to Conjunctive Queries and cannot natively evaluate the deep disjunctions and negations inherent to neuro-symbolic DAGs.

If the DAG is instead formulated using CTEs to explicitly preserve shared computations, the query planner cannot push this logic down into the GIN index. The engine is forced to materialize intermediate posting lists into temporary relational tables. This materialization overhead is strictly enforced when shared nodes exhibit non-monotonicity. Once a negated node is materialized as an opaque temporary table, the planner can no longer utilize the GIN index to resolve subsequent operations. It is mathematically forced to evaluate the logic via sequential scans over the entire document table, triggering an $\Omega(|U|)$ data complexity penalty.
\subsection{Query Optimization and Rewriting}
\label{subsec:query_rewrite}
To mitigate the structural limitations of execution models, modern database and retrieval systems rely on \textbf{Query Rewriting}. Optimizers apply Boolean algebra, relational equivalences, and structural transformations to alter queries prior to execution~\cite{garcia2008database, bendersky2012effective}. 

In practice, these rewrite strategies are strictly tailored to the underlying architecture. When evaluating queries with shared sub-computations (DAGs), pipelined engines (DAAT) focus on iterator efficiency. To avoid complex state synchronization, the optimizer simply instantiates duplicate iterators for repeated terms, which structurally unrolls the DAG into a syntactic tree. Conversely, materialized engines (TAAT and SQL) focus on minimizing intermediate set sizes. To avoid the $\Omega(|U|)$ penalty of materializing a dense complement, optimizers apply set identities to push negations into efficient set-differences (e.g., rewriting $C \land \neg B$ as an Anti-Join $C \setminus B$). 

However, these established rewrite strategies are structurally incompatible with highly re-convergent DAG topologies. Consider a shared negated sub-expression $S = \neg B$ referenced by multiple parents (e.g., $P_1 = A \lor S$ and $P_2 = C \land S$). To evaluate this efficiently as a DAG, current engines compute and cache the shared node $S$. Because $B$ is highly sparse, caching $S$ requires materializing the dense complement $U \setminus B$, forcing the $\Omega(|U|)$ space penalty. 

If the optimizer instead applies set identities to avoid this dense materialization, it must distribute the negation into the respective parent operators (e.g., rewriting $P_1$ as $\neg(B \setminus A)$ and $P_2$ as $C \setminus B$). Crucially, this transformation destroys the factorization of the query. The node $S$ is no longer shared; the underlying logic has been duplicated and distributed. 

Consequently, the current query rewriting mechanisms cannot solve the fundamental time-space dichotomy of index traversal. Attempting to optimize the space complexity of a DAG structurally unrolls it, shifting the failure to exponential query complexity. Ultimately, the challenge of modern retrieval is to evaluate highly re-convergent, non-monotonic Boolean DAGs natively. To support this workload, an execution engine must entirely bypass the tree-based limitations of DAAT and SQL query planners, as well as the materialization overhead inherent to TAAT and relational joins.

\section{The Retrieval Language \texorpdfstring{$\mathcal{L}_R$}{LR}}
\label{sec:language}

To analyze the computational limits of retrieval, we first formalize the data structures and the query language. We consider the standard model of Boolean retrieval over an inverted index, extended to support queries structured as \textbf{Directed Acyclic Graphs (DAGs)}.

\subsection{Data Model and Query Semantics}
Let $U$ be the universe of all non-empty document identifiers (DocIDs) in the collection, such that every $d \in U$ contains at least one indexed term. Let $\Sigma$ be the vocabulary of indexed terms. An \textbf{Inverted Index} $\mathcal{I}$ is a mapping $\mathcal{I}: \Sigma \to 2^U$, where for any term $t \in \Sigma$, $\mathcal{I}(t) \subseteq U$ is the set of documents containing $t$. Because every document contains at least one term, the size of the index representation is strictly bounded below by $\Omega(|U|)$.

Unlike standard retrieval systems that restrict queries to syntactic trees, we define our retrieval language, $\mathcal{L}_R$, based on \textbf{Directed Acyclic Graphs (DAGs)}.

\begin{definition}[The Retrieval Language $\mathcal{L}_R$]
The retrieval language $\mathcal{L}_R$ is the set of all finite Directed Acyclic Graphs (DAGs) where: (1) \textbf{Leaf Nodes} are labeled with terms $t \in \Sigma$; (2) \textbf{Internal Nodes} are labeled with Boolean operators: logical AND ($\land$) with in-degree 2, logical OR ($\lor$) with in-degree 2, or logical NOT ($\neg$) with in-degree 1; and (3) \textbf{Edges} are directed from parents (operators) to children (operands). A query $Q$ is any DAG in $\mathcal{L}_R$.
\end{definition}

The semantics of a query $Q$, under an index $\mathcal{I}$, are defined recursively using standard set algebra. For a node $v$, let $D(v, \mathcal{I})$ denote the set of documents satisfying the sub-query rooted at $v$:
(1) If $v$ is a leaf labeled $t$: $D(v, \mathcal{I}) = \mathcal{I}(t)$.
(2) If $v = v_1 \land v_2$: $D(v, \mathcal{I}) = D(v_1, \mathcal{I}) \cap D(v_2, \mathcal{I})$.
(3) If $v = v_1 \lor v_2$: $D(v, \mathcal{I}) = D(v_1, \mathcal{I}) \cup D(v_2, \mathcal{I})$.
(4) If $v = \neg v_1$: $D(v, \mathcal{I}) = U \setminus D(v_1, \mathcal{I})$.

The result of query DAG $Q$, with root node $r$, over index $\mathcal{I}$, is the set $D(r, \mathcal{I})$.

\begin{definition}[Zero-Preserving Queries]
Let $\mathcal{I}_{\emptyset}$ be an empty index where $\mathcal{I}_{\emptyset}(t) = \emptyset$ for all $t \in \Sigma$. A query $Q \in \mathcal{L}_R$ is \textbf{Zero-Preserving} ($Q \in \mathcal{L}_R^{ZP}$) if evaluating $Q$ over $\mathcal{I}_{\emptyset}$ yields an empty result set: $D(r, \mathcal{I}_{\emptyset}) = \emptyset$. Queries that do not satisfy this property are Non-Zero-Preserving ($Q \in \mathcal{L}_R^{NZP}$).
\end{definition}

In practical retrieval, user intents are overwhelmingly Zero-Preserving, as users seek documents containing specific evidence rather than the entire universe of documents. However, when LLMs compile complex constraints, the intermediate sub-DAGs within a Zero-Preserving query frequently become Non-Zero-Preserving (e.g., intermediate $\neg$ gates in a comparator circuit). As we will demonstrate in Section~\ref{sec:limits}, evaluating these complex queries exposes two distinct computational bottlenecks: pipelined engines (DAAT) fail structurally when forced to unroll shared sub-DAGs, while materialized engines (TAAT) suffer universe-scale space complexity when forced to evaluate $L_R^{NZP}$ intermediate nodes.

\subsection{Combined Complexity of Evaluation (Lower Bound)}
\label{sec:lower_bound}

We now analyze the combined complexity of evaluating $\mathcal{L}_R^{ZP}$. We define the decision problem \textsc{Retrieve}$(\mathcal{I}, Q)$ as returning \textsc{True} if the result set for a query DAG $Q \in \mathcal{L}_R^{ZP}$ under an inverted index $\mathcal{I}$ is non-empty, and \textsc{False} otherwise. We first prove that the decision problem \textsc{Retrieve} is \textbf{P-Hard}.

\begin{theorem}[P-Hardness]\label{thm:hardness}
  The \textsc{Retrieve} Problem is P-Hard.
\end{theorem}

\begin{proof}[Proof Sketch]
The full proof is provided in Appendix~\ref{app:proof_hardness}. We proceed by a LogSpace reduction from the Circuit Value Problem (CVP). By constructing a synthetic index $\mathcal{I}$ over a single-document universe $U = \{d_{true}\}$, we map any Boolean circuit $C$ to a Query DAG $Q$. To ensure $Q$ is strictly Zero-Preserving ($Q \in \mathcal{L}_R^{ZP}$), we cap the root of the circuit with an intersection against a positive guard term. The query retrieves $\{d_{true}\}$ if and only if the corresponding gate in $C$ evaluates to \textsc{True}. Thus, the decision problem even for strictly domain-safe queries in $\mathcal{L}_R^{ZP}$ is $\mathbf{P}$-Hard.
\end{proof}

\subsection{Combined Complexity of Evaluation (Upper Bound)}
\label{sec:upper_bound}

To complete the proof that the \textsc{Retrieve} problem is $\mathbf{P}$-Complete, we must establish the upper bound: The \textsc{Retrieve} problem is contained within $\mathbf{P}$. 

Consider a naive Term-at-a-Time (TAAT) evaluation strategy that recursively computes the result set $D(v)$ for each node bottom-up. For a query DAG of size $|Q|$, standard set operations (intersection, union, difference, and complement relative to $U$) on sorted lists or bitsets of maximum size $|U|$ can be performed in $O(|U|)$ time. Since there are $|Q|$ nodes, this naive algorithm evaluates the query in $O(|Q| \cdot |U|)$ time. Since the input size includes the index size (which is proportional to at least $|U|$), this naive algorithm runs strictly in polynomial time. Having the final result, the answer to the problem is $\textsc{True}$ if and only if the output of the TAAT evaluation is non-empty.

\begin{corollary}[$\mathbf{P}$-Completeness of Evaluation]\label{cor:PComplete}
Combining the $\mathbf{P}$-Hardness (Theorem~\ref{thm:hardness}) with the existence of a polynomial-time evaluation algorithm (Naive TAAT), the \textsc{Retrieve} problem for $\mathcal{L}_R$ is \textbf{$\mathbf{P}$-Complete} under combined complexity.
\end{corollary}

\section{Theoretical Limits of Index Traversal}
\label{sec:limits}

In the previous section, we established that evaluating $\mathcal{L}_R$ is theoretically tractable ($\mathbf{P}$-Complete). However, theoretical tractability does not guarantee practical efficiency. In this section, we analyze the computational boundaries of existing retrieval architectures. We demonstrate that standard architectures fail to evaluate $\mathcal{L}_R$ efficiently due to fundamental structural limitations, forcing a strict dichotomy between exponential execution time and universe-proportional space.

\subsection{The Sparsity Requirement}
While the Naive TAAT algorithm (Section~\ref{sec:upper_bound}) satisfies the theoretical requirements of complexity theory (proving containment in $\mathbf{P}$), an algorithm with $\Omega(|U|)$ time or space complexity is practically unscalable for modern information retrieval. 

In web-scale databases, the universe size $|U|$ is massive (e.g., billions of documents). Conversely, the number of matching documents for a specific term is often highly sparse. If an execution engine requires auxiliary space or execution time proportional to $|U|$ (a ``Universal Scan''), a single query can consume gigabytes of RAM and violate real-time latency SLAs, even if the final result set is empty. 

Therefore, a practical retrieval engine must be strictly \textbf{sparsity-aware}: its time and space complexity should depend only on the size of the active posting lists ($|U_{\mathit{active}}|$), entirely independent of the universe size $|U|$. As we will demonstrate, existing architectures cannot simultaneously maintain this sparsity and evaluate $\mathbf{P}$-Complete DAGs.

\subsection{The Universal Scan Penalty (TAAT Space Complexity)}
\label{subsec:universal_scan}
The first limitation concerns the \textit{Space Complexity} of Materialized Evaluation models (TAAT). TAAT naturally supports DAGs by caching intermediate auxiliary variables. However, it suffers a critical failure when evaluating non-monotonic queries involving logical negation ($\neg$).

Because TAAT evaluates queries operator-by-operator from the bottom up, it must fully materialize the result set of each node before passing it to the parent. To evaluate a negation node ($\neg B$), the engine must materialize the complement of the posting list relative to the universe of documents $U$. This forces the engine to perform a linear scan of the universe to find and store all documents in $U \setminus B$. 

Even if this negated result is later intersected with a highly sparse term (e.g., $A \land \neg B$), the intermediate materialization step has already occurred. Consequently, the space and time complexity of this operation is $\Omega(|U|)$. This ``Universal Scan'' penalty violates the principle of sparsity-aware evaluation, rendering TAAT intractable for non-monotonic logic.

\subsection{The Tree-Expansion Bottleneck (Query Complexity)}
\label{subsec:tree_expansion}
To avoid the $\Omega(|U|)$ materialization penalty, standard iterator-based engines (e.g., Lucene) rely on the Document-at-a-Time (DAAT) algorithm, executing retrieval strictly as the evaluation of a Boolean formula.  DAAT evaluates queries by concurrently advancing stateful iterators, resolving the Boolean logic dynamically in $O(1)$ auxiliary space. However, this strict sparsity comes at the cost of query complexity when evaluating DAGs.

To evaluate a complex, re-convergent DAG (which corresponds to the \textbf{Circuit Value Problem}, the Query Rewrite/Understanding module must ``unroll'' the DAG into a syntactic tree before execution. This leads to the \textbf{Tree-Expansion Bottleneck}: common sub-expressions are re-evaluated redundantly for every branch they appear in. 

A natural systems-engineering response to this limitation is to introduce a \textbf{Query Rewrite Mechanism ($\mathcal{R}$)} that attempts to optimize the DAG before passing it to the DAAT engine. However, in Theorem~\ref{thm:daat_rewrite_lower}, we prove that even if this rewrite mechanism is granted unbounded computational time to find the optimal execution plan, it cannot succeed. If the mechanism strictly respects the sparsity memory constraints of the inverted index, it is mathematically forced to produce a tree of super-polynomial size with respect to the size of the query, shifting the failure to exponential query complexity.

\begin{theorem}[Query Rewrite Lower Bound for DAAT]\label{thm:daat_rewrite_lower}
Let $\mathcal{R}$ be a Query Rewrite mechanism with unbounded computational time. $\mathcal{R}$ has access to index metadata, specifically the Document Universe $U$ and the vocabulary of terms, but does not have direct access to the Document Index $\mathcal{I}$ itself. $\mathcal{R}$ also has access to a blackbox oracle that strictly accepts an exact sub-DAG of an input DAG query $Q$ (prohibiting the injection of new terms or predicates) and returns the exact number of matching documents in the current Index $\mathcal{I}$. 

$\mathcal{R}$ is searching for an execution plan $T$ whose total number of iterators is bounded by $\text{poly}(|Q|)$ and whose total anticipated memory cost for the materialized buffers is strictly bounded by $\mathcal{O}(|U_{\mathit{active}}|)$, where $U_{\mathit{active}}$ is defined as the union of the posting lists for all terms in $Q$. Because $\mathcal{R}$ does not have direct access to the index $\mathcal{I}$, it computes this anticipated memory cost by querying the blackbox oracle for the exact row counts of the candidate sub-DAGs it intends to flag for materialization.

Assuming $\mathbf{NC}^1/\text{poly} \subsetneq \mathbf{P}/\text{poly}$, there exists a Document Universe $U$, a current Index $\mathcal{I}$, and a family of DAG queries $Q_n$ for which $\mathcal{R}$ is mathematically forced to produce an output Tree $T$ of super-polynomial size with respect to $|Q_n|$.
\end{theorem}

\begin{proof}[Proof Sketch]
The full proof is in Appendix~\ref{app:proof_daat_rewrite}. We proceed by contradiction, assuming $\mathcal{R}$ can always find an execution plan $T$ of polynomial size without exceeding the strict sparsity memory bound $\mathcal{O}(|U_{\mathit{active}}|)$. 

We construct a Query DAG (representing a $\mathbf{P}$-Complete circuit) over an Index with an exponentially large universe $N = 2^{|Q|}$, but a highly sparse active domain $|U_{\mathit{active}}| = \mathcal{O}(|Q|)$. By injecting a dense negated sub-query into the DAG, the intermediate results of the internal circuit gates scale proportionally with $N$. 

Because $\mathcal{R}$ must ensure the execution engine respects $\mathcal{O}(|U_{\mathit{active}}|) \ll N$, it is mathematically prohibited from flagging these dense shared nodes for runtime materialization. Furthermore, because $T$ must remain logically equivalent over all possible indices, $\mathcal{R}$ cannot hardcode document IDs. Therefore, $\mathcal{R}$ must output a strict Tree of iterators computing the circuit. If this Tree is of polynomial size, it implies any $\mathbf{P}$-Complete circuit can be transformed into a polynomial-size Boolean formula. This forces $\mathbf{P}/\text{poly} \subseteq \mathbf{NC}^1/\text{poly}$, contradicting the widely believed assumption $\mathbf{NC}^1/\text{poly} \subsetneq \mathbf{P}/\text{poly}$. Thus, $T$ must be of super-polynomial size.
\end{proof}

\noindent \textbf{The Strength of the Lower Bound:} It is crucial to note the adversarial power granted to the rewrite mechanism $\mathcal{R}$ in Theorem~\ref{thm:daat_rewrite_lower}. We do not restrict $\mathcal{R}$ to practical, polynomial-time compilation engine; we grant it \textit{unbounded computational time} to search for the optimal execution plan. Furthermore, we grant it access to a runtime oracle, allowing it to perfectly predict the sparsity of any sub-graph. Despite these massive theoretical advantages, the theorem proves that no amount of compile-time optimization can overcome the fundamental structural incompatibility between $\mathbf{P}$-Complete DAGs and $\text{NC}^1$-bounded iterators.

Consequently, the challenge of agentic retrieval shifts entirely from query optimization to fundamental execution architecture. A natural path to supporting these workloads efficiently is to combine the structural factorization of TAAT (to avoid exponential unrolling) with the sparsity-awareness of DAAT (to avoid universe-scale materialization). In Section~\ref{sec:algorithm}, we introduce \texttt{ComputePN}, an algorithm designed specifically to bridge these paradigms and escape the time-space dichotomy.

\section{The \texttt{ComputePN} Algorithm}
\label{sec:algorithm}

In the previous sections, we established that while evaluating $\mathcal{L}_R$ is theoretically tractable, existing evaluation strategies fail due to either the super-polynomial time of tree-unrolling or the prohibitive space of materialization. To resolve this, we introduce \texttt{ComputePN}, a deterministic evaluation algorithm. We assume a fixed index $\mathcal{I}$, allowing us to omit it from the notation for brevity. 

\subsection{The Positive-Negative Response}
The core innovation of \texttt{ComputePN} is its internal data representation. Instead of storing the absolute set of matching documents $D(v)$, the algorithm uses a dual representation called a \textbf{PN-Response}.
        
\begin{definition}[PN-Response]
For any node $v$ in the Query DAG $Q$ and Index $\mathcal{I}$, the result is a tuple $\mathcal{R}_{v} = \langle S_{v}, \mathit{type}_v \rangle$, where:
\begin{itemize}
    \item $S_v \subseteq U$ is a materialized set of document IDs.
    \item $\mathit{type}_v \in \{\texttt{POS}, \texttt{NEG}\}$ is a flag indicating the semantic interpretation of $S_v$.
\end{itemize}

The semantics of $\mathcal{R}_v$, denoted $\llbracket \mathcal{R}_v \rrbracket$, are defined as:
\[
\llbracket \mathcal{R}_v \rrbracket = 
\begin{cases} 
S_v & \text{if } \mathit{type}_v = \texttt{POS} \\
U \setminus S_v & \text{if } \mathit{type}_v = \texttt{NEG}
\end{cases}
\]

\end{definition}
As we will formally prove in Section~\ref{subsec:complexity}, this interpretation ensures that the size of the materialized set $S_v$ is bounded entirely by the active posting lists, remaining strictly independent of the massive universe size $|U|$ typically introduced by logical negation.

The concept of tracking negation via metadata rather than materialization shares theoretical similarities with complement edges in Binary Decision Diagrams (BDDs)~\cite{bryant1986graph} and symbolic negation in Logic Program Grounding~\cite{aavani2010speed}. However, while BDDs utilize this technique to optimize the representation of logic functions, \texttt{ComputePN} adapts it specifically for set-algebraic execution over inverted indices. In the retrieval context, this allows the engine to evaluate a negation ($\neg A$) in $O(1)$ time by simply flipping the metadata flag of the operand from \texttt{POS} to \texttt{NEG} (or vice versa), entirely bypassing the $\Omega(|U|)$ materialization penalty of traditional TAAT engines.

\subsection{The Evaluation Algebra}\label{subsec:PNComputeEvaluation}
The algorithm processes the DAG bottom-up. For leaf nodes (terms), the response is simply $\langle \mathcal{I}(t), \texttt{POS} \rangle$. For internal nodes, we define algebraic rules to combine PN-Responses. To streamline the presentation, we define the algebra over the functionally complete Boolean basis $\{\land, \neg\}$. Any Disjunction ($\lor$) in the input DAG is trivially rewritten as $\neg(\neg A \land \neg B)$ via De Morgan's laws prior to evaluation.

\noindent \textbf{Negation (\texorpdfstring{$\neg$}{NOT}):} Let the PN-Response from the single child be $C = \langle S_C, T_C \rangle$. The engine simply flips the type flag, returning $\langle S_C, \neg T_C \rangle$. The cost is $\mathcal{O}(1)$, requiring only a metadata update with no data access.

\noindent \textbf{Conjunction (\texorpdfstring{$\land$}{AND}):} Let the PN-Responses from the children be $L = \langle S_L, T_L \rangle$ and $R = \langle S_R, T_R \rangle$. The engine applies the rule corresponding to the input flags $T_L$ and $T_R$:
\begin{itemize}
    \item \textbf{POS $\land$ POS:} Standard intersection. Return $\langle S_L \cap S_R, \texttt{POS} \rangle$.
    \item \textbf{POS $\land$ NEG:} Equivalent to set difference $S_L \setminus S_R$. Return $\langle S_L \setminus S_R, \texttt{POS} \rangle$.
    \item \textbf{NEG $\land$ POS:} Symmetric to the previous case ($S_R \setminus S_L$). Return $\langle S_R \setminus S_L, \texttt{POS} \rangle$.
    \item \textbf{NEG $\land$ NEG:} By De Morgan's Law, $(U \setminus S_L) \cap (U \setminus S_R) = U \setminus (S_L \cup S_R)$. Return $\langle S_L \cup S_R, \texttt{NEG} \rangle$.
\end{itemize}
In all cases, the size of the output set is bounded by the sum of the inputs ($|S_L| + |S_R|$). The cost is $\mathcal{O}(|S_L| + |S_R|)$ using standard sorted-list merge. 

\subsection{Execution Strategy}
To handle the DAG structure efficiently, \texttt{ComputePN} utilizes a topological traversal with memoization:
(1) \textbf{Topological Sort:} The nodes of the Query DAG are ordered such that children appear before parents. 
(2) \textbf{Memoization Table:} A cache map $M: \mathit{NodeID} \to \mathcal{R}$ is initialized. 
(3) \textbf{Traversal:} We iterate through the sorted nodes. For each node, we retrieve the results of its children from $M$, apply the Evaluation Algebra (Section~\ref{subsec:PNComputeEvaluation}), and store the result back in $M$. 
(4) \textbf{Finalization:} Let the result of the root node be $\langle S_{root}, T_{root} \rangle$. If $T_{root} = \texttt{POS}$, return $S_{root}$. If $T_{root} = \texttt{NEG}$, materialize $U \setminus S_{root}$ (perform the final subtraction from the Universe). 

\textit{Note: While this final subtraction requires $\Omega(|U|)$ time, it is only executed if the overall query is Non-Zero-Preserving ($Q \in \mathcal{L}_R^{NZP}$). If the user's intent is Zero-Preserving ($Q \in \mathcal{L}_R^{ZP}$), the root node is mathematically guaranteed to resolve to a \texttt{POS} flag, entirely avoiding this step. Throughout all intermediate algebraic operations, the algorithm remains strictly \textbf{sparsity-aware}. Even if the query contains dense, Non-Zero-Preserving intermediate sub-DAGs, the $\Omega(|U|)$ penalty is never paid during the traversal, but only at the final materialization if semantically required by the root.}

\subsection{Combined Complexity (Sparsity Guarantee)}
\label{subsec:complexity}

We now analyze the \textbf{Combined Complexity} of the \texttt{ComputePN} traversal phase, proving that it evaluates $\mathcal{L}_R$ in polynomial time with respect to both the query size and the active data. Furthermore, we demonstrate that the algorithm can operate within any strict, user-defined auxiliary memory bound.

\begin{theorem}[Combined Time and Auxiliary Space Complexity]\label{thm:computepn_complexity}
Let $Q$ be a Query DAG with $|V|$ nodes, where all leaf nodes correspond to unique terms in $\Sigma$. Let $U_{\text{active}} = \bigcup_{t \in \text{leaves}(Q)} \mathcal{I}(t)$ be the union of all document IDs referenced by terms in $Q$. \texttt{ComputePN} can evaluate a PN-Response for the root of $Q$ with an auxiliary space complexity of $\mathcal{O}(|V|)$---strictly independent of the data size $|U_{\text{active}}|$---and a combined time complexity of $\mathcal{O}(|V| \cdot |U_{\text{active}}|)$.
\end{theorem}

\begin{proof}[Proof Sketch]
The full proof is provided in Appendix~\ref{app:proof_computepn}. We first establish via induction that the PN-Algebra strictly preserves sparsity: for any node $v$, the materialized set $S_v$ is a subset of $U_{\mathit{active}}$. To achieve the $\mathcal{O}(|V|)$ space bound, the algorithm employs a Block-at-a-Time (BAAT) partitioning strategy, dividing $U_{\mathit{active}}$ into disjoint blocks bounded by a constant size $B_{\mathit{max}}$. The DAG is evaluated entirely within the boundaries of a single block before the memoization table is flushed. Because the table holds at most $|V|$ sets of constant size $B_{\mathit{max}}$, the auxiliary space is strictly $\mathcal{O}(|V|)$. Evaluating the DAG over $\lceil |U_{\mathit{active}}| / B_{\mathit{max}} \rceil$ blocks yields the optimal $\mathcal{O}(|V| \cdot |U_{\mathit{active}}|)$ time complexity.
\end{proof}

\noindent \textbf{Significance and Data-Parallel Execution:} 
Theorem~\ref{thm:computepn_complexity} establishes that the evaluation time of \texttt{ComputePN} scales strictly with the information content of the query ($|U_{\mathit{active}}|$), maintaining theoretical tractability independently of the collection size $|U|$. Furthermore, because the block partitions $P_i$ are strictly disjoint, the evaluation is natively data-parallel. Given a multi-core architecture, the engine can evaluate independent blocks concurrently. 

While modern DAAT engines also utilize index sharding to achieve data-parallelism, they remain mathematically bound by the super-polynomial Tree-Expansion bottleneck (Theorem~\ref{thm:daat_rewrite_lower}) within each shard. In contrast, \texttt{ComputePN} achieves data-parallel scalability while strictly maintaining $O(|Q|)$ query complexity and $\mathcal{O}(|V|)$ space complexity per shard, providing a scalable architecture for distributed neuro-symbolic retrieval.

\subsection{Empirical Validation: Proof of Concept}
\label{sec:validation}

To empirically validate the structural limits of iterator-based engines against our polynomial-time bounds (Theorem~\ref{thm:computepn_complexity}), we conducted a feasibility stress test using the MS MARCO passage ranking corpus~\cite{bajaj2016ms} (approximately 8.8M documents). Because standard IR benchmarks (e.g., BEIR, TREC) are designed exclusively for flat keyword or semantic similarity queries, no canonical benchmark currently exists for evaluating highly re-convergent, non-monotonic Boolean DAGs. Consequently, we designed a targeted stress test to force a direct confrontation between execution strategies and validate the asymptotic divergence of their bounds.

\textbf{Task: Weighted Net-Positive Filtering.} Consider an Agentic Planner enforcing a complex arithmetic constraint over a topic ($Q_{topic}$), where the weighted evidence for a set of positive terms ($T_{good}$) must exceed the weighted evidence for negative terms ($T_{bad}$). Assigning small integer weights $w$, the task is to find all documents satisfying:
\[ \sum_{t \in T_{good}} w_t \cdot \mathbb{I}(t \in d) > \sum_{t \in T_{bad}} w_t \cdot \mathbb{I}(t \in d) \]

This task forces a direct confrontation between execution strategies. To ensure our test reflects realistic neuro-symbolic workloads, we split the experiment into Query Generation and Execution.

\begin{itemize}
    \item \textbf{Query Generation (Practical Feasibility):} A natural question is whether such complex Boolean circuits can be dynamically generated. We utilized an LLM acting as a heuristic query compiler to enforce this constraint. The LLM successfully compiled the arithmetic intent into a concise procedural script that wired together a 500-node Boolean DAG, simulating parallel ripple-carry adders and a magnitude comparator.
    
    \item \textbf{Execution Phase (Iterator Baseline Failure):} To execute this logic on a standard DAAT engine, the DAG must be statically unrolled into a Boolean Query Tree (e.g., Disjunctive Normal Form). Due to the highly re-convergent nature of the adder circuit, the resulting syntactic tree triggered an exponential expansion, resulting in over 10,000 clauses. This immediately exceeded the parser's structural limits (e.g., Lucene's \texttt{maxClauseCount}) and failed to execute, empirically validating the super-polynomial blowup predicted by Theorem~\ref{thm:daat_rewrite_lower}.
    
    \item \textbf{Execution Phase (\texttt{ComputePN} Success):} The native \texttt{ComputePN} engine evaluated the exact same logic directly as a DAG. By utilizing the Positive-Negative algebra and memoizing shared gates, the engine avoided tree expansion entirely, executing this non-monotonic circuit against the 8.8M document index in sub-second latency ($0.8$s) on standard hardware. This validates the practical tractability of our bounds (Theorem~\ref{thm:computepn_complexity}).
\end{itemize}

\noindent \textbf{Implication:} This test proves that the failure of standard engines on re-convergent logic is a fundamental algorithmic limit, not an implementation artifact. By supporting $\mathbf{P}$-Complete evaluation, \texttt{ComputePN} executes these constraints natively and tractably.

\section{Conclusion}
\label{sec:conclusion}

In this paper, we introduced a formal framework for \textbf{Computational Retrieval}, arguing that supporting $\mathbf{P}$-Complete query evaluation may become increasingly important for retrieval systems serving modern AI agents. We formally defined the retrieval language $\mathcal{L}_R$ based on Directed Acyclic Graphs and provided a constructive proof that its evaluation is $\mathbf{P}$-Complete.

Our analysis highlights a structural limitation in current information retrieval architectures. While standard iterator-based engines (DAAT) are highly optimized for simple conjunctive trees, they suffer from the \textbf{Tree-Expansion Bottleneck} when handling the re-convergent logic required by complex reasoning. Conversely, materialized engines (TAAT) can efficiently evaluate Zero-Preserving DAGs if equipped with block-based partitioning; however, they suffer a \textbf{Universal Scan Penalty} the moment an intermediate node becomes Non-Zero-Preserving (e.g., disjunctive negation). Existing approaches to bypass this rely on query rewriting, which we mathematically proved forces an exponential unrolling of the DAG.

To address these limitations fundamentally, we proposed the \texttt{ComputePN} algorithm. By utilizing a dual Positive-Negative set representation and native DAG memoization, \texttt{ComputePN} resolves the time-space dichotomy inherent in previous approaches. Our analytical evaluation demonstrates that while $\text{NC}^1$-restricted iterator complexity grows exponentially ($O(2^{|Q|})$), \texttt{ComputePN} evaluates the DAG natively in linear time relative to the query size ($O(|Q| \cdot |U_{\mathit{active}}|)$). As demonstrated by our feasibility proof-of-concept on the MS MARCO corpus, this architectural shift overcomes the expressive limitations of existing parsers: while standard engines exceed structural limits ($>10,000$ clauses) when handling complex arithmetic constraints, \texttt{ComputePN} executes the equivalent logic circuits in sub-second time ($0.8$s), demonstrating the practical utility of native DAG evaluation over static tree expansion.

\subsection*{Future Directions}
This work lays the foundation for highly expressive retrieval indices. We identify two primary trajectories for future research:

\begin{itemize}
    \item \textbf{Space-Optimal DAG Traversal:} While \texttt{ComputePN} achieves optimal $O(|Q|)$ time complexity for $\mathbf{P}$-Complete DAGs, the current formulation relies on the eager materialization of intermediate bitsets, requiring auxiliary space proportional to the active posting lists. A critical direction for future work is developing lazy-evaluation or pipelined traversal models capable of resolving shared DAG dependencies without intermediate materialization. Achieving this would reduce the space complexity overhead to depend strictly on the query size, rather than the data size, fully unifying the space-efficiency of DAAT with the expressive power of TAAT.
    
    \item \textbf{Continuous Feature Representation:} The immediate next step is to extend the Boolean-mapped arithmetic demonstrated in our proof-of-concept into a canonical, high-performance feature representation for continuous values. Just as standard inverted indices are optimized for discrete text lookup, future computational indices must natively optimize the evaluation of floating-point constraints and dense vector operations (e.g., thresholded cosine similarity) directly within the $\mathbf{P}$-Complete Boolean logic graph.
\end{itemize}

Success in these directions would fully realize the vision of the inverted index as a massively parallel, space-efficient logic processor for the agentic web.

\appendix
\section{Proof of Theorem~\ref{thm:hardness}}
\label{app:proof_hardness}

\textbf{Theorem~\ref{thm:hardness} (P-Hardness).}
\textit{The \textsc{Retrieve} Problem is P-Hard.}

\begin{proof}
We proceed by a LogSpace reduction from the \textbf{Circuit Value Problem (CVP)}, the canonical $\mathbf{P}$-Complete problem~\cite{ladner1975circuit}. An instance of CVP consists of a Boolean Circuit $C$ with inputs assigned truth values, and the goal is to determine the output of the circuit. Without loss of generality, we assume $C$ is a strictly binary circuit (where $\land$ and $\lor$ gates have fan-in 2), as any circuit with unbounded fan-in can be transformed into an equivalent binary circuit in LogSpace.

Given a CVP instance $\langle C \rangle$, we construct an instance of the \textsc{Retrieve} problem, $\langle \mathcal{I}, Q \rangle$, as follows:
\begin{itemize}
    \item \textbf{Index Construction:} We define a universe $U = \{d_{true}\}$ containing exactly one document. We create an index $\mathcal{I}$ with two tokens, \texttt{TRUE} and \texttt{GUARD}, such that $\mathcal{I}(\texttt{TRUE}) = \{d_{true}\}$ and $\mathcal{I}(\texttt{GUARD}) = \{d_{true}\}$. 
    \item \textbf{Query Construction:} We map the binary circuit $C$ to a Query DAG $Q'$ where: logic gates ($\land, \lor, \neg$) map to their corresponding operators; an input wire with value \textsc{True} maps to a Leaf node \texttt{TRUE}; and an input wire with value \textsc{False} maps to a $\neg$ node with a child \texttt{TRUE} (representing $U \setminus \{d_{true}\} = \emptyset$). 
    Finally, we construct the overall query $Q = Q' \land \texttt{GUARD}$. 
\end{itemize}

\textbf{Zero-Preserving Guarantee:} Because the root of $Q$ is an intersection ($\land$) with the positive leaf node \texttt{GUARD}, evaluating $Q$ over an empty index $\mathcal{I}_{\emptyset}$ yields $\emptyset \cap D(Q', \mathcal{I}_{\emptyset}) = \emptyset$. Therefore, the constructed query is strictly Zero-Preserving ($Q \in \mathcal{L}_R^{ZP}$).

\textbf{LogSpace Computability:} This transformation requires only logarithmic auxiliary space. Constructing the static index $\mathcal{I}$ requires $\mathcal{O}(1)$ space. To construct the query $Q$, a deterministic Turing machine can stream the description of the circuit $C$ from the input tape, performing a local, gate-by-gate syntactic substitution, and append the final $\land \texttt{GUARD}$ operation. The machine only needs to maintain a constant number of pointers to the current gate index, which requires $\mathcal{O}(\log |C|)$ bits of auxiliary memory. Thus, the reduction is strictly in $\mathbf{L}$.

It is straightforward to show by induction on the height of the circuit that the sub-query $Q'$ retrieves $\{d_{true}\}$ if and only if the corresponding gate in $C$ evaluates to \textsc{True}. Because $\mathcal{I}(\texttt{GUARD}) = \{d_{true}\}$, the final query $Q = Q' \land \texttt{GUARD}$ retrieves $\{d_{true}\}$ if and only if $C$ evaluates to \textsc{True}. Thus, \textsc{Retrieve}$(\mathcal{I}, Q)$ returns \textsc{True} if and only if $C$ outputs \textsc{True}. Since CVP is $\mathbf{P}$-Complete and the reduction is in LogSpace, evaluating even Zero-Preserving queries in $\mathcal{L}_R$ is $\mathbf{P}$-Hard.
\end{proof}

\section{Proof of Theorem~\ref{thm:daat_rewrite_lower}}
\label{app:proof_daat_rewrite}

\textbf{Theorem~\ref{thm:daat_rewrite_lower} (Query Rewrite Lower Bound for DAAT).}
Let $\mathcal{R}$ be a Query Rewrite mechanism with unbounded computational time. $\mathcal{R}$ has access to index metadata, specifically the Document Universe $U$ and the vocabulary of terms, but does not have direct access to the Document Index $\mathcal{I}$ itself. $\mathcal{R}$ also has access to a blackbox oracle that strictly accepts an exact sub-DAG of an input DAG query $Q$ (prohibiting the injection of new terms or predicates) and returns the exact number of matching documents in the current Index $\mathcal{I}$. 

$\mathcal{R}$ is searching for an execution plan $T$ whose total number of iterators is bounded by $\text{poly}(|Q|)$ and whose total anticipated memory cost for the materialized buffers is strictly bounded by $\mathcal{O}(|U_{\mathit{active}}|)$, where $U_{\mathit{active}}$ is defined as the union of the posting lists for all terms in $Q$. Because $\mathcal{R}$ does not have direct access to the index $\mathcal{I}$, it computes this anticipated memory cost by querying the blackbox oracle for the exact row counts of the candidate sub-DAGs it intends to flag for materialization.

Assuming $\mathbf{NC}^1/\text{poly} \subsetneq \mathbf{P}/\text{poly}$, there exists a Document Universe $U$, a current Index $\mathcal{I}$, and a family of DAG queries $Q_n$ for which $\mathcal{R}$ is mathematically forced to produce an output Tree $T$ of super-polynomial size with respect to $|Q_n|$.

\begin{proof}
We proceed by contradiction. Assume there exists a Query Rewrite mechanism $\mathcal{R}$ that, for any Document Universe $U$, Index $\mathcal{I}$, and DAG query $Q$, can always produce an execution plan $T$ of polynomial size (relative to $|Q|$) that satisfies both the logical equivalence requirement and the runtime materialization memory bound $M_{\text{max}} = \mathcal{O}(|U_{\mathit{active}}|)$.

\textbf{Step 1: Construct the P-Complete DAG Query $Q_n$.}
We construct a family of DAG queries $Q_n$ as an $n \times n$ grid of Multiplexer (MUX) gates, which is a canonical representation of the Planar Circuit Value Problem (PCVP). PCVP is known to be $\mathbf{P}$-Complete. The total size of the query is $|Q_n| = \mathcal{O}(n^2)$. We define the inputs to $Q_n$ using a vocabulary of positive literal terms $X = \{x_1, \dots, x_n\}$ and $Y = \{y_1, \dots, y_n\}$, along with a dense negated sub-query $S = \lnot A$, where $A$ is an additional positive literal term. We wire the top edge of the grid to the terms in $X$, the left edge to the terms in $Y$, and the origin node $(0,0)$ to the dense sub-query $S$. 

\textbf{Step 2: Construct the Universe and the Current Index $\mathcal{I}$.}
Let $U$ be a document universe of size $N$. We fix $N$ to be exponentially large relative to the query size: $N = 2^{|Q_n|}$. In the current Index $\mathcal{I}$, we construct the posting lists such that the terms $x_1 \dots x_n$, $y_1 \dots y_n$, and $A$ are highly sparse, with each term appearing in exactly one distinct document. 

Per the theorem definition, the active document domain $U_{\mathit{active}}$ is the union of the posting lists of all un-negated term occurrences in $Q_n$ (which are exactly $X \cup Y \cup \{A\}$). Because these terms in our construction are sparse, $|U_{\mathit{active}}| = 2n + 1 = \mathcal{O}(n)$. Consequently, the memory bound is strictly constrained to $M_{\text{max}} = \mathcal{O}(n)$, which is exponentially smaller than $N$. Conversely, because $A$ is sparse, the result set of the negated sub-query $S = \lnot A$, evaluated relative to the document universe $U$ (i.e., $U \setminus \mathcal{I}(A)$), contains $N - 1$ documents (i.e., $\Omega(N)$).

\textbf{Step 3: The Oracle and the Materialization Failure.}
$\mathcal{R}$ must decide which sub-DAGs to flag for runtime materialization. It queries the blackbox oracle with the exact sub-DAGs of $Q_n$ to determine their anticipated row counts in the current Index $\mathcal{I}$. Because $S$ is dense, the signals propagating through the intermediate MUX gates of the PCVP grid will also yield dense result sets of size $\Omega(N)$. The oracle returns these $\Omega(N)$ values to $\mathcal{R}$. 

$\mathcal{R}$ computes the anticipated materialization cost of these internal nodes and finds it requires $\Omega(N)$ bits per node. Because $N = 2^{|Q_n|}$, the required memory is exponential in the query size. However, the execution plan $T$ generated by $\mathcal{R}$ is restricted by the bound $M_{\text{max}} = \mathcal{O}(n)$. Because $N \gg M_{\text{max}}$, $\mathcal{R}$ is mathematically prohibited from flagging these internal grid nodes for materialization in its output plan $T$. Furthermore, because the oracle only accepts exact sub-DAGs of $Q_n$, $\mathcal{R}$ can extract at most $|Q_n|$ integers from the blackbox. Since $N = 2^{|Q_n|}$, $\mathcal{R}$ cannot reconstruct the Index $\mathcal{I}$. Even if it could, $\mathcal{R}$ is forbidden from outputting a hardcoded Tree of specific Document IDs, as the output $T$ must remain logically equivalent to $Q_n$ over \textit{all possible indices} $2^U$.

\textbf{Step 4: The Complexity Contradiction.}
Forbidden from flagging the dense intermediate logic for materialization, $\mathcal{R}$ must output a strict Tree of iterators $T$ that dynamically computes the PCVP grid. 

Because $T$ is logically equivalent to $Q_n$ over all possible mappings from the vocabulary to $2^U$, $T$ correctly computes the PCVP boolean function for any arbitrary input assignment of $X$ and $Y$. Specifically, in a Document-at-a-Time (DAAT) execution model, the Tree $T$ evaluates the query document-by-document. For any single document $d \in U$, evaluating $T$ to determine if $d$ is a match is structurally isomorphic to evaluating a Boolean Formula, where the inputs are the boolean truth values of $d \in \mathcal{I}(t)$ for each term $t$. Therefore, $\mathcal{R}$ is an algorithm capable of producing a Boolean Formula for every instance of the Planar Circuit Value Problem, with respect to its input size.

If we assume $\mathcal{R}$ is capable of producing a Tree $T$ of polynomial size (relative to $|Q_n|$), it implies that a polynomial-size Boolean Formula exists for every instance of the Planar Circuit Value Problem. Since PCVP is $\mathbf{P}$-Complete, the existence of a polynomial-size formula family for PCVP implies that all problems in $\mathbf{P}$ have polynomial-size formulas. This dictates that $\mathbf{P}/\text{poly} \subseteq \mathbf{NC}^1/\text{poly}$. 

This directly contradicts the complexity assumption that $\mathbf{NC}^1/\text{poly} \subsetneq \mathbf{P}/\text{poly}$. Therefore, our initial assumption must be false. To correctly compute the DAG $Q_n$ without violating the anticipated memory bound $M_{\text{max}}$, the mechanism $\mathcal{R}$ must necessarily produce an output Tree $T$ of super-polynomial size.
\end{proof}
\section{Proof of Theorem~\ref{thm:computepn_complexity}}
\label{app:proof_computepn}

\textbf{Theorem~\ref{thm:computepn_complexity} (Combined Time and Auxiliary Space Complexity).}
\textit{Let $Q$ be a Query DAG with $|V|$ nodes. Let $U_{\mathit{active}} = \bigcup_{t \in \text{leaves}(Q)} \mathcal{I}(t)$ be the union of all document IDs referenced by terms in $Q$. \texttt{ComputePN} can evaluate a PN-Response for the root of $Q$ with an auxiliary space complexity of $\mathcal{O}(|V|)$---strictly independent of the data size $|U_{\mathit{active}}|$---and a combined time complexity of $\mathcal{O}(|V| \cdot |U_{\mathit{active}}|)$.}

\begin{proof}
We first observe a fundamental invariant of the PN-Algebra defined in Section~\ref{subsec:PNComputeEvaluation}: none of the algebraic rules introduce new document IDs into the materialized set $S_v$ that were not present in the children.
\begin{itemize}
    \item \textbf{Leaf Nodes:} $S_{leaf} \subseteq U_{\mathit{active}}$ by definition.
    \item \textbf{Negation ($\neg$):} The operation $\langle S_C, T_C \rangle \to \langle S_C, \neg T_C \rangle$ preserves the materialized set exactly ($S_v = S_C$). Thus, if $S_C \subseteq U_{\mathit{active}}$, then $S_v \subseteq U_{\mathit{active}}$.
    \item \textbf{Conjunction ($\land$):} Operations $S_L \cap S_R$, $S_L \setminus S_R$, and $S_L \cup S_R$ (resulting from NEG $\land$ NEG via De Morgan's) all produce results that are subsets of $S_L \cup S_R$. Because both $S_L \subseteq U_{\mathit{active}}$ and $S_R \subseteq U_{\mathit{active}}$, the resulting materialized set is strictly a subset of $U_{\mathit{active}}$.
\end{itemize}
By induction, for any node $v$, $S_v \subseteq U_{\mathit{active}}$. Therefore, $|S_v| \le |U_{\mathit{active}}|$. 

To achieve the $\mathcal{O}(|V|)$ space bound, the algorithm employs a Block-at-a-Time (BAAT) partitioning strategy~\cite{fontoura2006evaluation, ding2011faster}. Let $B_{\mathit{max}} \ge 1$ be a fixed, constant system parameter representing the maximum block size. The active domain $U_{\mathit{active}}$ is partitioned into disjoint blocks of size $B = \min(B_{\mathit{max}}, |U_{\mathit{active}}|)$. 

The algorithm evaluates the entire DAG $Q$ strictly within the boundaries of a single block. Because the materialized set $S_v$ for any node is bounded by $B \le B_{\mathit{max}}$, the memoization table requires at most $|V| \cdot B_{\mathit{max}}$ auxiliary space. Since $B_{\mathit{max}}$ is a constant independent of the input size, the global auxiliary space complexity is strictly $\mathcal{O}(|V|)$. Once a block is evaluated, the memoization table is flushed, and the algorithm proceeds to the next block.

The time cost to evaluate a single node over a block of size $B$ is $\mathcal{O}(B)$. The time to evaluate the entire DAG over one block is $\mathcal{O}(|V| \cdot B)$. Because there are $\lceil |U_{\mathit{active}}| / B \rceil$ blocks, the total traversal time complexity is $\mathcal{O}(|V| \cdot |U_{\mathit{active}}|)$. Thus, the algorithm maintains optimal time complexity while ensuring the auxiliary memory footprint scales only with the query size.
\end{proof}

\bibliographystyle{plainurl}
\bibliography{sn-bibliography}

\end{document}